\documentclass{article}

\usepackage{arxiv}
\usepackage[utf8]{inputenc} 
\usepackage[T1]{fontenc}    
\usepackage{hyperref}       
\usepackage{url}            
\usepackage{booktabs}       
\usepackage{amsfonts}       
\usepackage{nicefrac}       
\usepackage{microtype}      
\usepackage{graphicx}
\usepackage{doi}
\usepackage{amsmath}
\usepackage{authblk}

\title{Adaptive Frequency Bin Interval in FFT via Dense Sampling Factor $\alpha$}

\author{ \href{https://orcid.org/0009-0008-1100-4580}{\includegraphics[scale=0.06]{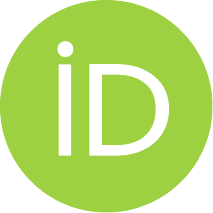}\hspace{1mm}Haichao Xu} \\
	Institute for Astronomy, School of Physics, Zhejiang University\\
	866 Yuhangtang Rd, Hangzhou, 310058, China \\
	\href{Haichao_XU@zju.edu.cn}{Haichao\_XU@zju.edu.cn} \\
}

\date{}

\renewcommand{\headeright}{preprint}
\renewcommand{\undertitle}{preprint}

\begin{document}
\maketitle

\begin{abstract}
	The Fast Fourier Transform (FFT) is a fundamental tool for signal analysis, widely used across various fields. However, traditional FFT methods encounter challenges in adjusting the frequency bin interval, which may impede accurate spectral analysis. In this study, we propose a method for adjusting the frequency bin interval in FFT by introducing a parameter $\alpha$. We elucidate the underlying principles of the proposed method and discuss its potential applications across various contexts. Our findings suggest that the proposed method offers a promising approach to overcome the limitations of traditional FFT methods  and enhance spectral analysis accuracy.
\end{abstract}


\section{Introduction}

After the pioneering work of Cooley and Tukey in developing the Fast Fourier Transform (FFT) algorithm \cite{cooley1965algorithm}, the Discrete Fourier Transform (DFT) has become indispensable in various fields for signal analysis\cite{oppenheim1999discrete,cooleyFastFourierTransform1969}. 
However, traditional FFT methods are plagued by the picket fence effect (PFE), which constrains the resolution of discrete frequency bins. The spacing of discrete frequencies (known as bin interval) and the number of calculation points are restricted in traditional FFT methods, impeding the effective characterization of spectral features at desired frequencies.

Two techniques have been developed to mitigate the PFE: bin interpolation and zero-padding. In the 1970s, the method of bin interpolation was incorporated into the FFT, giving rise to the creation of the Interpolated FFT approach \cite{jainHighAccuracyAnalogMeasurements1979,rifeUseDiscreteFourier1970}. This technique interpolates between adjacent frequency spectrum data points to estimate spectral features at frequencies not directly sampled. However, bin interpolation inherently involves spectral estimation, which can introduce significant bias, particularly under coherence conditions or in the presence of high noise levels. In contrast, zero-padding involves the addition of extra zeros to the end of the signal sequence to reduce the intervals between frequency bins. This method enhances the utilization of the signal sequence and enables customizable bin intervals based on requirements. However, zero-padding does not modify the core FFT algorithm, thereby introducing additional computational overhead, especially when aiming for very small bin intervals.

In this paper, we propose a novel approach to alter the bin interval in the DFT. Building upon the conventional techniques, our method introduces a flexible mechanism to adjust the frequency bin interval, enhancing the versatility of spectral analysis. Inspired by the structure of the FFT, we present an accelerated form of this method, aiming to improve computational efficiency while preserving accuracy. 
This innovative approach not only expands the capabilities of traditional DFT methods but also streamlines the spectral analysis process, making it more adaptable to diverse signal processing tasks.

\section{A Method to Enhance Flexibility of Bin Interval in DFT}

To facilitate our discussion, let's start with a signal $x(t)$ containing noise. Its Fourier transform and inverse transform are given by:

\begin{equation}
  \left\{\begin{aligned}
    &X(\nu)=\int_{-\infty}^{\infty}e^{-2\pi i \nu t}x(t)dt\\
    &x(t)=\int_{-\infty}^{\infty}e^{2\pi i \nu t}X(\nu)d\nu
  \end{aligned}\right..
  \label{eq:continuous_Fourier}
\end{equation}

However, continuous recording of such a signal is often impractical. Instead, we sample the signal at intervals $\Delta t$ within a finite time interval $[0, T]$, obtaining a signal sequence $\{x_n\}$ of length $N$. Similarly, we compute a finite number of spectral data points. To capture a broader frequency range, we set the upper frequency limit as twice the Nyquist frequency, acknowledging that higher frequencies correspond to negative frequencies. Employing a matrix representation of the Fourier transform, we transform the time-domain signal sequence into an equally sized frequency-domain signal sequence. Thus, the discretized form of (\ref{eq:continuous_Fourier}) is expressed as:

\begin{equation}
  \left\{\begin{aligned}
    &X_m=\frac{T}{N}\sum_{n=0}^{N-1}\left[\exp\left(-2\pi i\frac{mn}{N}\right)x_n\right]\\
    &x_n=\frac{1}{T}\sum_{m=0}^{N-1}\left[\exp\left(2\pi i\frac{mn}{N}\right)X_m\right]
  \end{aligned}\right..
  \label{eq:discrete_Fourier}
\end{equation}

Comparing (\ref{eq:continuous_Fourier}) and (\ref{eq:discrete_Fourier}), we observe that the DFT essentially involves the following substitutions:

\begin{equation}
  t \rightarrow \frac{n}{N}T, \quad dt \rightarrow \frac{T}{N}, \quad \nu \rightarrow \frac{m}{T}, \quad d\nu \rightarrow \frac{1}{T}, \quad m, n = 0, 1, \cdots, N-1.
  \label{eq:substitute_DFT}
\end{equation}

To ensure that the spectral signal returns to the original time-domain signal upon inverse Fourier transformation, we need the orthogonality normalization among different harmonics in the Fourier transform. In the discrete form (\ref{eq:discrete_Fourier}), the orthogonality normalization of harmonics is expressed as:

\begin{equation}
  \frac{1}{N}\sum_{m=0}^{N-1}\exp\left(2\pi i\frac{m(n-l)}{N}\right)=\sum_{k=-\infty}^{\infty}\delta_{n}^{l+kN},
  \label{eq:original_norm}
\end{equation}
in which $\delta_{m}^{n}$ denotes the Kronecker delta symbol, defined as:

\begin{equation*}
  \delta_{m}^{n}=\left\{
    \begin{aligned}
      1, & \quad \text{if } m=n \\
      0, & \quad \text{if } m \neq n
    \end{aligned}
  \right..
\end{equation*}

We incorporate the time duration $T$ from (\ref{eq:discrete_Fourier}) into $X_m$ and adjust the position of $1/N$ to conform with the DFT formulation in other literature, i.e.

\begin{equation}
  \left\{\begin{aligned}
    &X_m=\sum_{n=0}^{N-1}\left[\exp\left(-2\pi i\frac{mn}{N}\right)x_n\right]\\
    &x_n=\frac{1}{N}\sum_{m=0}^{N-1}\left[\exp\left(2\pi i\frac{mn}{N}\right)X_m\right]
  \end{aligned}\right..
  \label{eq:original_DFT}
\end{equation}

\begin{figure}[htb]
  \centering
  \includegraphics[width=0.50\textwidth]{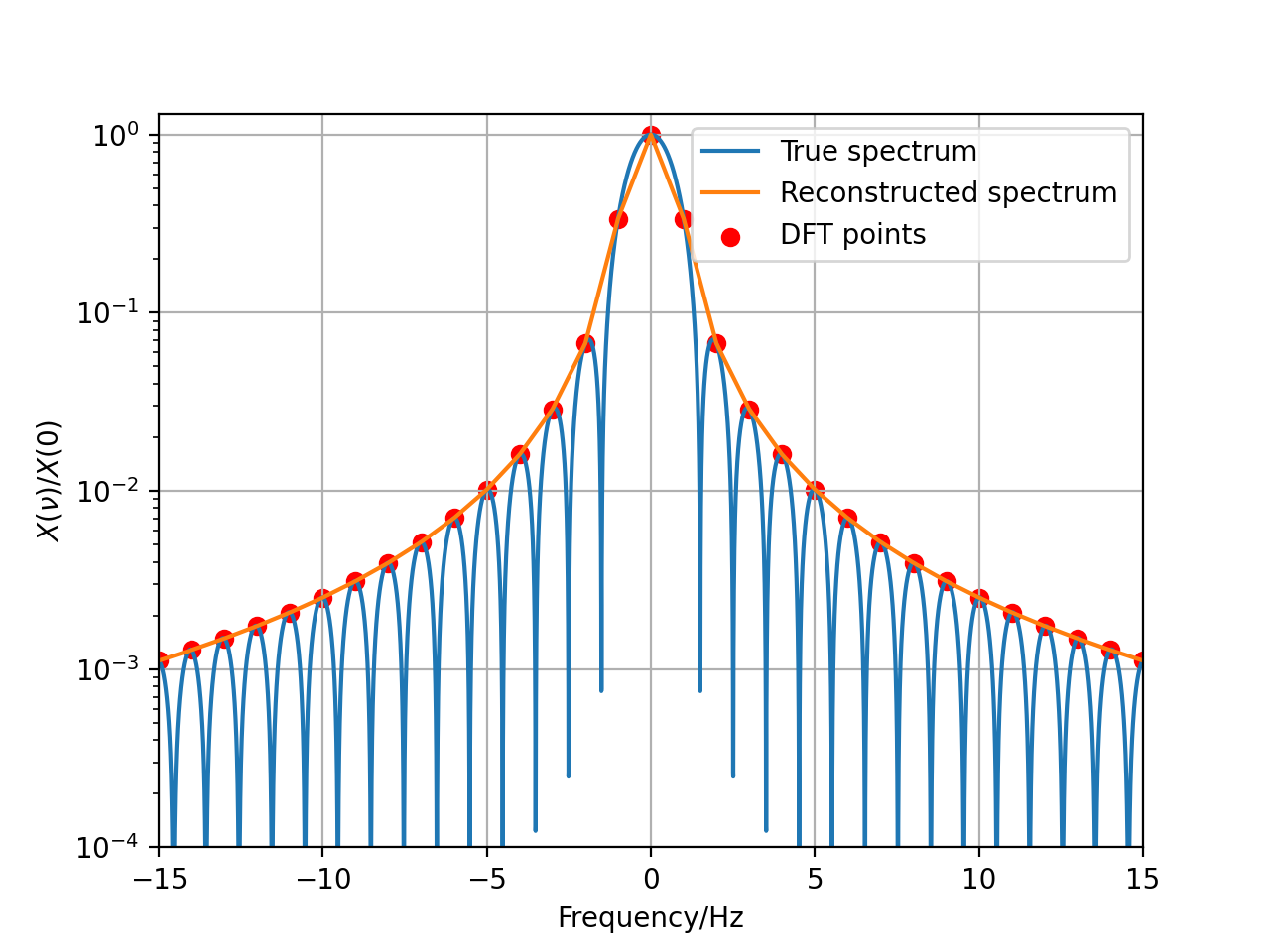} 
  \caption{Comparison between the reconstructed spectrum of DFT and the true spectrum of the signal $\sin(\pi t)$ defined over the interval $(0, 1)$}
  \label{fig:Limitation_DFT}
\end{figure}

The spectrum computed through equation (\ref{eq:original_DFT}) reveals that due to the fixed spacing in frequency space, spectra are only attainable at specific frequencies. As indicated by (\ref{eq:substitute_DFT}), maintaining a constant sampling time implies that even if we reduce the sampling interval, the frequency bin interval remains unaffected. Consider a signal $x(t) = \sin(\pi t)$ sampled over $[0,1]$. We compute its spectrum using both (\ref{eq:original_DFT}) and (\ref{eq:continuous_Fourier}). To reconcile discrepancies in Fourier transform coefficients, we normalize the $X(0)$ obtained from both methods individually. From Fig. (\ref{fig:Limitation_DFT}), we observe that without zero-padding, it is challenging to reconstruct accurate spectral features from densely sampled spectral points.

Now we aim to adjust the frequency values within the DFT framework while preserving the frequency range from 0 to twice the Nyquist frequency, which is necessary for subsequent FFT acceleration. Without changing the time-domain signal, we introduce a dense sampling factor $\alpha$ to modify (\ref{eq:substitute_DFT}) as follows:
\begin{equation}
  \begin{aligned}
    &t \rightarrow \frac{n}{N}T, ~ dt \rightarrow \frac{T}{N}, ~~ n = 0, 1, \cdots, N-1 ,\\
    &\nu \rightarrow \frac{m}{\alpha T}, ~ d\nu \rightarrow \frac{1}{\alpha T}, ~~ m = 0, 1, \cdots, \alpha N-1 .
  \end{aligned}
  \label{eq:substitute}
\end{equation}

Subsequently, we adapt (\ref{eq:original_DFT}) and (\ref{eq:original_norm}) to:
\begin{equation}
  \left\{
  \begin{aligned}
    &X_m = \sum_{n=0}^{N-1} \left[ \exp \left( -2\pi i \frac{mn}{\alpha N} \right) x_n \right] \\
    &x_n = \frac{1}{\alpha N} \sum_{m=0}^{\alpha N-1} \left[ \exp \left( 2\pi i \frac{mn}{\alpha N} \right) X_m \right]
  \end{aligned}
  \right.
  \label{eq:DFT}
\end{equation}
and
\begin{equation}
  \frac{1}{\alpha N} \sum_{m=0}^{\alpha N-1} \exp \left( 2\pi i \frac{m(n-l)}{\alpha N} \right) = \sum_{k=-\infty}^{\infty} \delta_{n}^{l+k\alpha N}.
  \label{eq:norm}
\end{equation}

\begin{figure}[htb]
\centering
\includegraphics[width=0.50\textwidth]{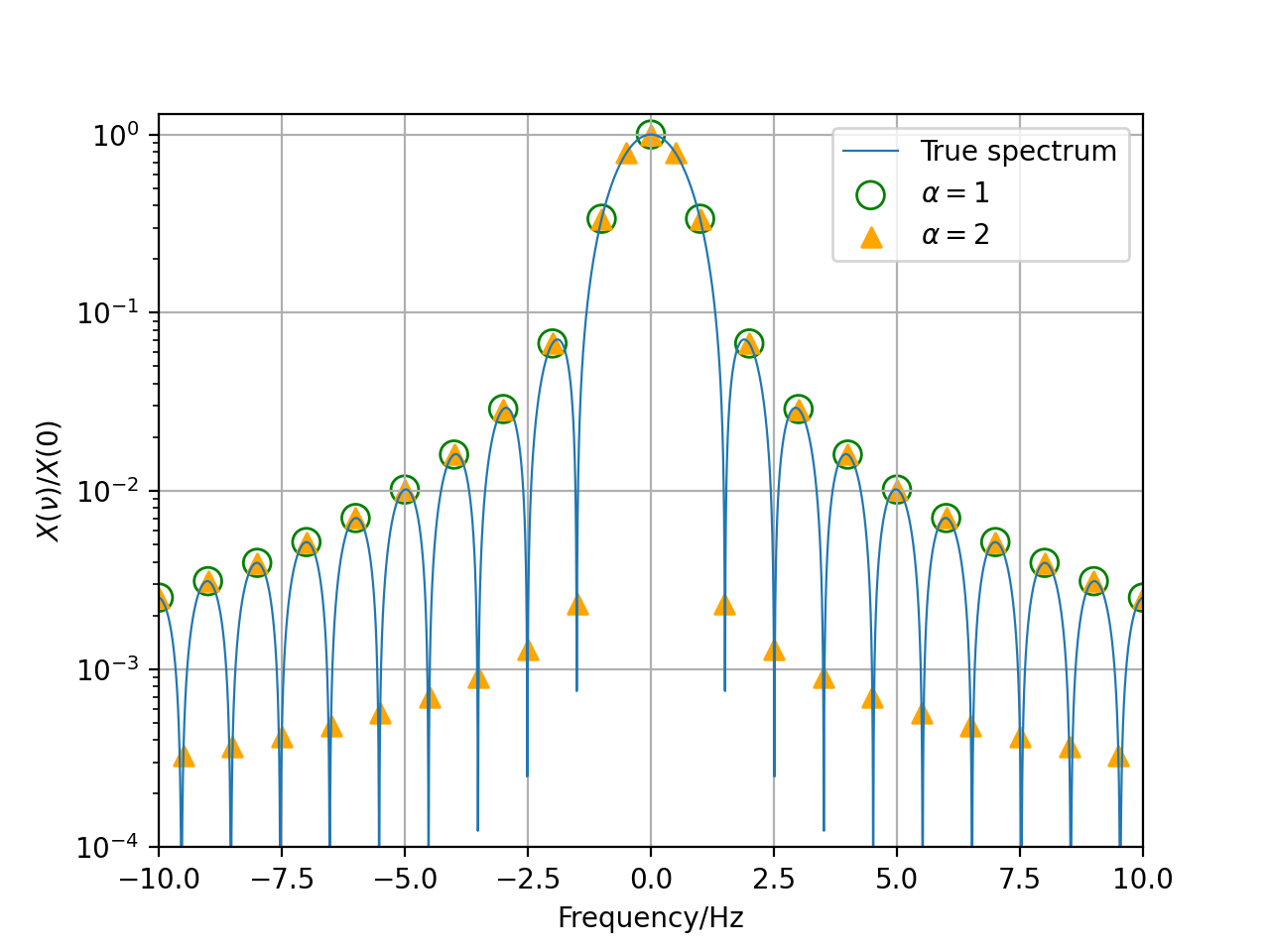}
\caption{DFT spectra points of $\sin(\pi t)$ for different $\alpha$ values}
\label{fig:DFT_highresolution}
\end{figure}

To achieve a smaller frequency bin interval for an existing time-domain signal sequence, we can set $\alpha$ greater than 1, according to (\ref{eq:substitute}). In this scenario, (\ref{eq:DFT}) is mathematically equivalent to appending $(\alpha-1)N$ zeros to the end of the signal sequence. We omit the additional proof here. As depicted in Fig. (\ref{fig:DFT_highresolution}), for the same signal sequence, increasing the value of $\alpha$ can better reflect the true spectral characteristics, especially when spectral amplitude changes significantly at different frequencies.

Conversely, if the data quality is high and spectral changes are gradual, the conventional DFT method becomes computationally intensive due to the computation of numerous spectral points. In such cases, to alleviate this computational burden, $\alpha$ can be set to less than 1, although this leads to an increased frequency bin interval. However, when $\alpha<1$, the rank of the DFT matrix is given by the row rank $\alpha N$, indicating substantial information loss during the DFT process. Consequently, $\alpha<1$ results in a one-way Fourier transform, making inverse transformation impractical for recovering the original time-domain signal. If we continue to utilize (\ref{eq:DFT}) for computation, the reconstructed time-domain signal $x'_n$ based on (\ref{eq:norm}) will become:
\begin{equation*}
x'_n = \sum_{k} x_{n+k\alpha N},
\end{equation*}
where the summation encompasses all $0\leq n+k\alpha N<N$.

\section{Accelerating Spectral Analysis with FFT Algorithm}

To enhance the practical applicability of the adjustable bin interval DFT method in signal analysis, it is crucial to explore an acceleration approach akin to the classical FFT algorithm. In the following derivation, we will utilize the general FFT algorithm developed by Cooley and Tukey\cite{cooleyFastFourierTransform1969}, employing divide-and-conquer techniques and recursion, which can be further expedited using the butterfly diagram method. For convenience, we adopt the widely accepted convention:

\begin{equation*}
  W_{N}^{mn}=\exp\left(-2\pi i\frac{mn}{N}\right).
\end{equation*}

In (\ref{eq:DFT}), the DFT transformation corresponds to an $N \times \alpha N$ matrix operation. Examining the coefficients $W_{\alpha N}^{mn}$ in (\ref{eq:DFT}), we identify three key properties, in which the appearance of 2 arises from our adoption of splitting the problem into two smaller subproblems:

\begin{equation}
  \left\{
  \begin{aligned}
    & W_{\alpha N}^{m(n+1)}=W_{\alpha N}^{m}W_{\alpha N}^{mn}\\
    & W_{\alpha N}^{2mn}=W_{\alpha N/2}^{mn}\\
    & W_{\alpha N}^{m+\frac{\alpha N}{2}}=-W_{\alpha N}^{m}
  \end{aligned}
  \right..
  \label{eq:property}
\end{equation}

We can employ a similar strategy to the classical FFT algorithm by splitting the DFT into computations for even and odd indices of the signal sequence:

\begin{equation}
  X_l=\sum_{k=0}^{N/2-1}\left(W_{\alpha N}^{l\cdot(2k)}x_{2k}\right)+\sum_{k=0}^{N/2-1}\left(W_{\alpha N}^{l\cdot(2k+1)}x_{2k+1}\right).
  \label{eq:divide}
\end{equation}

We can introduce two sequences of length $N/2$, denoted as $y_k$ and $z_k$, representing the components of even and odd indices in the signal sequence, respectively:
\begin{equation}
  y_k=x_{2k},~~z_k=x_{2k+1}.
\end{equation}
According to (\ref{eq:DFT}), the DFT transformations of these sequences are given by:
\begin{equation}
  \left\{
  \begin{aligned}
    & Y_l=\sum_{k=0}^{N/2-1}\left(W_{\frac{\alpha N}{2}}^{lk}y_k\right)\\
    & Z_l=\sum_{k=0}^{N/2-1}\left(W_{\frac{\alpha N}{2}}^{lk}z_k\right)
  \end{aligned}
  \right.,
\end{equation}
both of which correspond to $\frac{N}{2} \times \frac{\alpha N}{2}$ matrix operations.

Since the general divide-and-conquer process is widely discussed in literature, such as \cite{oppenheim1999discrete,press2007numerical}, we will refrain from delving into excessive detail here. By applying the three properties in (\ref{eq:property}) sequentially, we can simplify (\ref{eq:divide}) as follows:
\begin{equation}
  \left\{\begin{aligned}
    X_l &= Y_l + W_{\alpha N}^{l}Z_l \\
    X_{l+\frac{\alpha N}{2}} &= Y_l - W_{\alpha N}^{l}Z_l
  \end{aligned}\right..
\end{equation}
This processing effectively reduces an $N \times \alpha N$ matrix operation to two separate $\frac{N}{2} \times \frac{\alpha N}{2}$ matrix operations while introducing two additional linear combinations. This allows the divide operation to proceed recursively by preserving the matrices' shape while only adjusting their sizes.

In contrast to the classical FFT operations, our computation involves Fourier matrices that are not necessarily square. Consequently, the termination criterion for recursion differs from that of the classical FFT. Each division operation preserves the shape of the Fourier matrix while reducing its scale. Eventually, the transformation matrix is reduced to rank 1 after several division steps, indicating the impracticality of further computation. For $\alpha > 1$, the matrix shape eventually becomes $\alpha \times 1$, terminating recursion when the signal sequence length reduces to 1. Conversely, for $\alpha < 1$, the matrix shape becomes $1 \times \frac{1}{\alpha}$, leading to recursion termination with
\begin{equation*}
  X_0 = \sum_{k=0}^{1/\alpha} x_k.
\end{equation*}
Alternatively, we can express the the number of divide-and-conquer operations with a unified formula:
\begin{equation}
  \mathcal{N} = \min \left\{ N, \alpha N \right\}.
\end{equation}
Similarly, let $\mathcal{M} = \max \left\{ N, \alpha N \right\}$. Introducing $\alpha$ and utilizing FFT algorithm acceleration, the computational complexity is approximated by:
\begin{equation}
  T_{\alpha}(n) =  \mathcal{O} \left( \mathcal{M} \log(\mathcal{N}) \right).
  \label{eq:complexity}
\end{equation}

\section{Discussion}

The method introduced in the article offer significant flexibility across various scenarios. People can customize the appropriate value of $\alpha$ according to different needs without changing the original signal sequence. 

When aiming to enhance spectral resolution, we can employ the zero-padding method by appending $(\alpha-1)N$ zeros to the end of the time-domain signal. The computational complexity of the conventional FFT operation with zero-padding is approximately $T_{p+F}(n) \approx \mathcal{O}(\alpha N \log(\alpha N)).$ In contrast, the method proposed in this article exhibits a computational complexity of $\mathcal{O}(\alpha N \log(N))$, according to (\ref{eq:complexity}), resulting in a difference in computational load between the two approaches:
\begin{equation}
  \Delta T(n) = T_{p+F}(n) - T_{\alpha}(n) \approx \mathcal{O}(\alpha N \log(\alpha)).
\end{equation}

In scenarios requiring high resolution, characterized by a considerably large $\alpha$, the method proposed in this article offers substantial computational savings compared to the conventional FFT with zero-padding combined approach. This is due to the redundancy of computation introduced by extra zeros after several recursive steps. When the signal sequence in the recursion operation becomes $\{x_i,0,0,\cdots\}$, further division is unnecessary.

Similarly, when the inverse transformation to retrieve the time-domain signal is unnecessary and spectral variations in the frequency domain are insignificant, conserving computational resources becomes feasible by employing $\alpha$ values less than 1. It's important to note that in order to accurately display the spectrum characteristics, a sufficient number of points in the spectrum must be maintained, so $\alpha N \gg 1$ is needed. In such cases, the computational savings achieved by the method in this article compared to the conventional FFT algorithm are approximated by:
\begin{equation}
  \Delta T(n) = T_F(n) - T_{\alpha}(n) \approx \mathcal{O}( N \log(\frac{1}{\alpha})).
\end{equation}

\section{Conclusion}

In this paper, we introduce a novel method for adjusting the frequency bin interval in the FFT with a dense sampling factor $\alpha$, which is efficient and scalable for practical applications.  Researchers can tailor the frequency bin interval according to specific signal characteristics and analysis requirements.  This adaptability expands the capabilities of spectral analysis techniques and streamlines the analysis process, making it more adaptable to diverse signal processing tasks.  Overall, our findings suggest that the proposed method has significant potential to advance signal analysis techniques across various domains. Further optimization of the method for specific use cases is also expected.

\bibliographystyle{ieeetr}
\bibliography{reference.bib}  

\begin{thebibliography}{1}

\bibitem{cooley1965algorithm}
J.~W. Cooley and J.~W. Tukey, ``An algorithm for the machine calculation of
  complex {{Fourier}} series,'' vol.~19, no.~90, pp.~297--301.

\bibitem{oppenheim1999discrete}
A.~V. Oppenheim, {\em Discrete-Time Signal Processing}.
\newblock Pearson Education India.

\bibitem{cooleyFastFourierTransform1969}
J.~W. Cooley, P.~A.~W. Lewis, and P.~D. Welch, ``The {{Fast Fourier Transform}}
  and {{Its Applications}},'' vol.~12, no.~1, pp.~27--34.

\bibitem{jainHighAccuracyAnalogMeasurements1979}
V.~K. Jain, W.~L. Collins, and D.~C. Davis, ``High-{{Accuracy Analog
  Measurements}} via {{Interpolated FFT}},'' vol.~28, no.~2, pp.~113--122.

\bibitem{rifeUseDiscreteFourier1970}
D.~C. Rife and G.~A. Vincent, ``Use of the discrete fourier transform in the
  measurement of frequencies and levels of tones,'' vol.~49, no.~2,
  pp.~197--228.

\bibitem{press2007numerical}
W.~H. Press, {\em Numerical Recipes 3rd Edition: {{The}} Art of Scientific
  Computing}.
\newblock Cambridge university press.

\end{thebibliography}






\end{document}